\documentclass[preprint,aps,showkeys,nofootinbib,12pt]{revtex4-1}
\usepackage{amsmath,amssymb,amsfonts,dcolumn,color,graphicx,graphics,latexsym,placeins,epsfig}
\usepackage{epsfig}
\usepackage{bm}
\usepackage{slashed}
\usepackage{latexsym}
\usepackage{natbib}
\usepackage{url}
\usepackage{dcolumn}
\usepackage{xcolor}
\usepackage{amsfonts,amssymb,amsmath}
\usepackage{graphicx,epsfig}
\usepackage{psfrag}
\usepackage{subfigure}
\usepackage{tabularx}
\usepackage{hyperref}
\usepackage[normalem]{ulem}

\allowdisplaybreaks

\newcommand{\be}{\begin{equation}}
\newcommand{\ee}{\end{equation}}
\newcommand{\ba}{\begin{eqnarray}}
\newcommand{\ea}{\end{eqnarray}}

\newcommand{\p}{\prime}

\newcommand{\del}{\partial}
\newcommand{\tdelta}{\tilde{\nabla}}
\newcommand{\nn}{\nonumber}

\newcommand{\hatg}{\hat{g}}

\def \dd {\mathrm{d}}
\def \t {\tilde}
\def \p {\partial}

\begin{document}

\title{Horndeski theories and beyond from higher dimensions}

\author{Soumya Jana$^{1,2}$\footnote{Soumya.Jana@etu.unige.ch, soumyajana.physics@gmail.com}, Charles Dalang$^{1}$\footnote{charles.dalang@unige.ch} and Lucas Lombriser$^{1}$\footnote{lucas.lombriser@unige.ch}}

\affiliation{$^{1}${\it D\'epartement de Physique Th\'eorique, Universit\'e de Gen\`eve, 24 quai Ernest Ansermet, 1211 Gen\`eve 4, Switzerland}, }

\affiliation{$^{2}${\it Department of Physics, Sitananda College, Nandigram, 721631, India }}


\begin{abstract}
The Einstein-Hilbert action with a cosmological constant is the most general local four-dimensional action leading to second-order derivative equations of motion that are symmetric and divergence free.
In higher dimensions, additional terms can appear. We investigate a generalised metric decomposition involving a scalar degree of freedom to express the higher-dimensional action as an effective four-dimensional scalar-tensor theory.
From the higher-dimensional Ricci scalar alone and a subclass of our metric ansatz, we recover the subset of Horndeski theories with luminal speed of gravitational waves.
More generally, beyond-Horndeski terms appear.
When including a Gauss-Bonnet scalar in the higher-dimensional action, we generate contributions to all cubic-order second-derivative terms present in the degenerate higher-order scalar-tensor theory as well as higher-derivative terms beyond that.
We discuss this technique as a way to generate healthy four-dimensional gravity theories with an extra scalar degree of freedom and outline further generalisations of our method.
\end{abstract}
\keywords{modified gravity, scalar-tensor theories, Horndeski theories, extra dimensions, cosmology}


\maketitle
\tableofcontents

\section{Introduction }

After more than a century, the theory of general relativity remains the most successful description of gravity.
From the deflection of light around massive bodies, lunar-laser ranging, Shapiro time-delay constraints, probes of the equivalence principle to gravitational waves, general relativity passes all tests to date while the parameter space for allowed deviations shrinks as time passes by~\cite{Will:2014kxa}.

The field equations of general relativity can be obtained from an elegant least-action principle and appropriate boundary conditions.
The Ricci scalar and a cosmological constant constitute the geometric sector of the Lagrangian.
Their variation with respect to the metric provides the geometric part of the field equations, a linear combination of the Einstein and metric tensors.
The metric variation of the matter sector of the action defines the energy-momentum tensor, the material part of the Einstein field equations.
Covariant energy-momentum conservation requires this contribution to be divergence free, which is also satisfied by the geometric side due to the Bianchi identities.
It is tempting to ask if there are other possibilities for the left-hand side of the Einstein field equations. 
What is the most general symmetric and divergence-free tensor that can be built from the metric and its first and second derivatives? As it turns out from the Lanczos-Lovelock theorem~\cite{Lanczos,Lovelock}, in four dimensions, the Einstein tensor is the only symmetric and divergence-free tensor of rank two which does not contain more than second-order derivatives of the metric tensor. This places general relativity in a very special place within the ocean of conceivable gravity theories.

 Nevertheless, general relativity is not free of problems. In the ultra-violet regime, one expects a breakdown of the theory which is non-renormalisable and as such loses its predictive power. On the other end of the energy spectrum, the infrared behaviour of general relativity
causes for surprises.
If one assumes general relativity to hold, then one needs to invoke the existence of an extensive dark sector, which so far has escaped Earth-based laboratories.
Even if one relaxes this assumption, the last decades have shown that modifications of gravity~\cite{Clifton:2011jh,Berti:2015itd,Joyce:2016vqv} motivated as alternatives to either dark matter or the late-time accelerated cosmic expansion are both severely challenged by observations~\cite{Nieuwenhuizen:2016uxv,Lombriser:2016yzn}.
Cosmic acceleration is in principle well explained by the cosmological constant and general relativity, but theoretical calculations of its value attributed to vacuum fluctuations of the matter fields exceed by many orders of magnitude the observed one, and any fine-tuning is furthermore radiatively unstable~\cite{Martin:2012bt} (see, however, Refs.~\cite{Barrow:2010xt,Kaloper:2013zca,Wang:2017oiy,Lombriser:2019jia,Sobral-Blanco:2020rdu}).
To go beyond general relativity, one must break one of the assumptions going into building the Einstein-Hilbert action.
Different possibilities include breaking Lorentz invariance, abandon locality, adding new degrees of freedom, or extra spacetime dimensions.
The connection between the last two options shall be the interest of this work.

The minimal choice in adding new degrees of freedom is the extension of gravity with one single degree of freedom, a scalar field.
Hereby, the Horndeski action~\cite{Horndeski:1974wa} defines the most general Lorentz-invariant 4-dimensional local metric theory of gravity\footnote{It is the analogue to general relativity for three propagating degrees of freedom. General relativity is contained as a subset.} that supplements general relativity by one scalar degree of freedom with restriction to second-order equations of motion.
The condition of second-order equations allows to evade Ostrogradsky ghost instabilities~\cite{Ostrogradsky1850}.
While this is a sufficient condition, it is not strictly necessary, as was realised in Ref.~\cite{deRham:2011qq}.
Beyond-Horndeski terms evading the ghosts were found in Ref.~\cite{Gleyzes:2014dya} and finally, the quest of the most general scalar-tensor theory led to the development of degenerate higher-order scalar-tensor theory (DHOST)~\cite{Motohashi:2016ftl,BenAchour:2016fzp}, which are explicitly known up to third order in second derivatives of the fields. The degeneracy condition for arbitrary higher-derivative theories were established in Refs.~\cite{Motohashi:2017eya,Motohashi:2018pxg}.

An {\it a priori} completely different path is to extend gravity to higher spacetime dimensions. 
In $D>4$ dimensions, the Einstein-Hilbert action is no longer the most general action that leads to divergence-free symmetric tensors in the second-order field equations and other terms may arise~\cite{Lanczos,Lovelock,Padmanabhan:2013xyr}.
%
%
%
It is well known that for an adequate metric decomposition involving a scalar field these extra dimensions can be integrated out to produce an effective scalar-tensor theory~\cite{Kaluza,Klein,Charmousis:2014mia,vandeBruck:2018jlz}. An interesting question that arises is whether from the most general theory of gravity with sufficiently many co-dimensions and the most general metric ansatz with a scalar field one recovers the most general scalar-tensor theory. 

The paper is organised as follows. In Sec.~\ref{sec:Einstein-Hilbert}, we start from the higher-dimensional Einstein-Hilbert action and show that a general metric decomposition involving a scalar field but being independent of its kinetic term leads to the full subset of Horndeski theories with luminal speed of gravitational waves.
A more general ansatz with kinetic dependency introduces additional, beyond-Horndeski terms.
Next, in Sec.~\ref{sec:Gauss-Bonnet}, we generalise the action with the inclusion of the higher-dimensional Gauss-Bonnet term and show that it is sufficient to generate contributions to all known DHOST terms after integrating out the extra dimensions while also generating new extra higher-order derivative terms beyond DHOST. In Sec.~\ref{sec:generalisations}, we discuss how further generalisations of our method will lead to even more general scalar-tensor and other extended gravity theories.
We summarise our results and conclude in Sec.~\ref{sec:conclusions}.
Finally, we provide further technical details of our derivations and definitions in the appendices A and B.

\section{$D$-dimensional Einstein-Hilbert action} \label{sec:Einstein-Hilbert}

We first consider the Einstein-Hilbert action with a cosmological constant in $D=4+d$ dimensions,
\begin{equation}
S=\frac{M^2_{4+d}}{2}\int \sqrt{-\det(g_{AB})} \left[R- 2\Lambda_{4+d}\right]\dd^{4+d}x \,,
\label{eq:EH_4+d}
\end{equation}
where $d$ indicates the number of co-dimensions, $g_{AB}$ is the $D$-dimensional metric with capital indices $A, B \in \{0, 1, \dots, D-1\}$, $R$ denotes the $D$-dimensional Ricci scalar and $M^2_{4+d}$ and $\Lambda_{4+d}$ are the squared Planck mass and the cosmological constant with appropriate dimensions, respectively.
For the spacetime line element we consider a general decomposition, which for the sake of simplicity, is block diagonal,
\begin{equation}
\dd s^2=g_{AB}\dd x^A\dd x^B= \tilde{g}_{\alpha \beta}(x^\mu)\dd x^{\alpha}\dd x^{\beta}+ \gamma_{mn}(\phi(x^{\mu}),X(x^{\mu}),y^i)\dd y^m \dd y^n\,,
\label{eq:decomposition}
\end{equation} 
where $X=-\frac{1}{2}\tilde{g}^{\alpha \beta} \del_{\alpha}\phi \del_{\beta}\phi$ is the kinetic term of the scalar field $\phi$, Greek indices satisfy $(\alpha, \beta, \mu)= \lbrace 0,1,2,3 \rbrace$ and Latin indices $(m,n,i)=\lbrace 3+1, 3+2,....(3+d) \rbrace$ run in the co-dimensions.
Tensors built from the effective $4$-dimensional metric $\tilde{g}_{\alpha\beta}$ exclusively are noted with a tilde.
The metric $\gamma_{mn}$ applies to the submanifold of the extra spatial dimensions. It is clear that this choice of block diagonal metric simplifies much of the calculations by avoiding the complicated calculation of a determinant in $D$ dimensions. With this choice, assuming $\gamma_{mn}$ to be a function of $\phi(x^{\alpha})$, $X(x^{\alpha})$, and the extra dimensions $y$ is quite general. 

The dimensionality reduction scheme is quite straightforward. We express the Lagrangian in Eq.~\eqref{eq:EH_4+d} as tensors which depend on the four spacetime dimensions, while the leftovers depend on the co-dimensions and only indirectly on the four spacetime dimensions through $\phi(x^\alpha)$ and $X(x^\alpha)$. In this way, the integral over the extra dimensions can be performed over the functions of the scalar field and its derivative, which matches the structure of general scalar-tensor theories.

\subsection{Tensor decomposition}

We start by computing the Christoffel symbols. From Eq.~\eqref{eq:decomposition}, we get
\begin{eqnarray}
\Gamma^{\alpha}_{\mu\nu}&=&\tilde{\Gamma}^{\alpha}_{\mu\nu}(x^{\beta})\,,\label{eq:gammabeg}\\
\Gamma^{\alpha}_{ab}&=& - \frac{1}{2}\gamma_{ab,\phi}\,\phi^{\alpha} + \frac{1}{2}\gamma_{ab,X}\, \phi^{\alpha}{}_{\beta}\phi^{\beta}\,,\\
\Gamma^{a}_{bc}&=& \bar{\Gamma}^{a}_{bc}(\phi,X,y)\,, \\
\Gamma^{a}_{\alpha b}&=& \frac{1}{2} \gamma^{ac}\left[\gamma_{bc,\phi} \phi_{\alpha}- \gamma_{bc,X}\phi_{\alpha\beta}\phi^{\beta} \right]\,,\\
\Gamma^{\alpha}_{\mu a}&=&\Gamma^a_{\mu\nu}=0 \label{eq:gammaend}\,,
\end{eqnarray}
where $\Gamma$, $\tilde{\Gamma}$, and $\bar{\Gamma}$ are the Christoffel symbols for the $D$-dimensional spacetime corresponding to the metric $g_{AB}$, the effective 4-dimensional spacetime corresponding to $\tilde{g}_{\alpha\beta}$, and the submanifold of extra spatial dimensions corresponding to $\gamma_{mn}$, respectively. We use the notations $\phi_{\alpha}\equiv \p_\alpha \phi= \tdelta_{\alpha}\phi$, $\phi_{\alpha\beta}\equiv \tdelta_{\alpha}\tdelta_{\beta}\phi$, $A_{,\phi}\equiv \frac{\partial A }{\partial \phi}$, and $A_{,X}\equiv\frac{\partial A}{\partial X}$ for any tensor $A$.

The next step is to obtain the elements of the Riemann tensor. Using Eqs.~\eqref{eq:gammabeg}--\eqref{eq:gammaend}, we obtain
\begin{eqnarray}
R^{\alpha}{}_{\beta \mu \nu}&=& \tilde{R}^{\alpha}{}_{\beta \mu \nu}.\\
R^{\alpha}{}_{\beta a b}&=& A_{ab}\left[\phi_{\beta\sigma}\phi^{\alpha}\phi^{\sigma}-\phi^{\alpha}{}_{\sigma}\phi_{\beta}\phi^{\sigma} \right],\\
R^{\alpha}{}_{a\beta b}&=& - \frac{1}{2}\gamma_{ab,\phi}\phi_{\beta}{}^{\alpha}+ B_{ab}\phi_{\beta}\phi^{\alpha} + C_{ab} \phi^{\alpha}\phi_{\beta\rho}\phi^{\rho} \\
& &+ \bar{C}_{ab}\phi_{\beta}\phi^{\alpha}{}_{\rho}\phi^{\rho} + D_{ab} \phi^{\alpha}{}_{\mu}\phi_{\beta\nu}\phi^{\mu}\phi^{\nu}- \frac{1}{2} \gamma_{ab,X}X_{\beta}{}^{\alpha}\,, \\
R^{\alpha}{}_{abc}&=& \phi^{\alpha}E_{abc} + \phi^{\alpha}{}_{\beta}\phi^{\beta}\bar{E}_{abc}\,,\\
R^a{}_{bcd}&=& \bar{R}^a{}_{bcd}-2XF^a{}_{bcd}-\phi^{\mu}\phi_{\mu\nu}\phi^{\nu} H^a{}_{bcd} + \phi^{\mu}\phi_{\mu\alpha}\phi^{\alpha\nu}\phi_{\nu} J^a{}_{bcd} \,, \\
R^a{}_{bc\alpha}&=& \phi_{\alpha} K^a{}_{bc} +\phi_{\alpha\beta}\phi^{\beta} \bar{K}^a{}_{bc}\,,\\
R^a{}_{b\alpha \beta}&=& A^a{}_{b}\left(\phi_{\beta\mu}\phi^{\mu}\phi_{\alpha}-\phi_{\alpha\mu}\phi^{\mu}\phi_{\beta}\right) \,, \\
R^a{}_{\alpha\mu\nu}&=&0\,, \\
R^{\alpha}{}_{a\mu\nu}&=&-\gamma_{ab}\tilde{g}^{\alpha\beta}R^b{}_{\beta\mu\nu}=0\,,\\
R^{a}{}_{\alpha b\beta}&=& \gamma^{ac}\tilde{g}_{\alpha\mu} R^{\mu}{}_{c\beta b} \,,\\
R^a{}_{\alpha b c}&=& -\gamma^{ad}\tilde{g}_{\alpha\beta}R^{\beta}{}_{dbc} \,, \\
 R^{\alpha}{}_{\beta\mu a}&=&0 \,,
\end{eqnarray}
where all the rank two, three, and four tensors on the right-hand side depend on the extra dimensional metric $\gamma_{mn}$ and its derivatives. Their definitions are given in Appendix \ref{app:extratensors}. 
The non-vanishing Ricci tensors are:
\begin{eqnarray}
R_{\alpha \beta}&=& R^A{}_{\alpha A \beta}= \tilde{R}^{\rho}{}_{\alpha \rho \beta}+ R^a{}_{\alpha a \beta}\nonumber\\
&=& \tilde{R}_{\alpha \beta} -\frac{1}{2} \gamma_\phi \phi_{\beta\alpha} + B \phi_{\beta}\phi_{\alpha} + C \phi_{\alpha} \phi^{\mu}\phi_{\beta \mu} +  \bar{C} \phi_{\beta} \phi^{\mu}\phi_{\alpha \mu}+ D \phi_{\alpha \mu}\phi_{\beta\nu}\phi^{\mu}\phi^{\nu} -\frac{1}{2}\gamma_X X_{\beta\alpha} \,, \\
R_{ab}&=&  R^A{}_{a A b}= R^{\alpha}{}_{a\alpha b} + R^c{}_{a c b}\nonumber\\
&=& \bar{R}_{ab}-\frac{1}{2}\gamma_{ab,\phi} \tilde{\square} \phi -2X\left(B_{ab}+F_{ab}\right) +  \left(C_{ab}+\bar{C}_{ab}-H_{ab}\right)\phi^{\mu}\phi_{\mu\nu}\phi^{\nu}\nonumber \\
&& + (D_{ab}+J_{ab})\phi^{\mu}\phi_{\mu\alpha}\phi^{\alpha\nu}\phi_{\nu}-\frac{1}{2}\gamma_{ab,X} \tilde{\square} X \,, \\
R_{a\alpha }&=& R^A{}_{a A \alpha}= R^{\beta}{}_{a \beta \alpha} + R^b{}_{ab\alpha}\nonumber\\
&=& \phi_{\alpha}K_a + \bar{K}_a \phi_{\alpha\beta}\phi^{\beta} \,,  \\
R_{\alpha a} &=& R^A_{\alpha A a}= R^{\beta}{}_{\alpha \beta a} + R^b{}_{\alpha b a}\nonumber\\
&=& -\phi_{\alpha} \gamma^{bd}E_{dba}-\phi_{\alpha\beta}\phi^{\beta}\gamma^{bd}\bar{E}_{dba} \nonumber\\
&=& \phi_{\alpha}K_a + \bar{K}_a \phi_{\alpha\beta}\phi^{\beta},
\end{eqnarray}
where
\begin{eqnarray}
\gamma_{\phi}=& \gamma^{ab}\gamma_{ba,\phi}\,, \qquad B= \gamma^{ab}B_{ba}\,, \qquad C= \gamma^{ab}C_{ba}\,, \\ \gamma_{X}=& \gamma^{ab}\gamma_{ba,X} \,,  \qquad D= \gamma^{ab}D_{ba}\,, \qquad \bar{C}=\gamma^{ab}\bar{C}_{ba}\,,
\end{eqnarray}
and $F_{ab}=F^c{}_{a c b}$, $H_{ab}=H^c{}_{a c b}$, $J_{ab}=J^c{}_{a c b}$, $K_a=K^b{}_{ab}$ and $\bar{K}_a=\bar{K}^b{}_{ab}$ are traced with the co-dimensional metric $\gamma^{ab}$. The Ricci scalar in $4+d$ dimensions can be expressed as
\begin{eqnarray}
R &=& \tilde{g}^{\alpha\beta}R_{\alpha\beta}+\gamma^{ab}R_{ab}\nonumber\\
&=& \tilde{R}+ \bar{R}_d-\gamma_{\phi} \tilde{\square} \phi - 2X(2B + F)+(2C+2\bar{C}-H)\phi^{\mu}\phi_{\mu\nu}\phi^{\nu}\nonumber\\
&& + (2D + J)\phi^{\mu}\phi_{\mu\alpha}\phi^{\alpha\nu}\phi_{\nu} -\gamma_X \tilde{\square}X \,,
\label{eq:ricci_4+d}
\end{eqnarray}
where $\tilde{R}$ is the Ricci scalar of the usual 4-dimensional spacetime and $\bar{R}_d=\gamma^{ab}\bar{R}_{ab}$ is the Ricci scalar of the submanifold of the extra $d$ dimensions.
The remaining terms depend on the metric $\gamma$, the scalar field and its kinetic term. We have defined $F\equiv\gamma^{ab}F_{ab}$, $H \equiv \gamma^{ab}H_{ab}$, and $J\equiv \gamma^{ab}J_{ab}$.

\subsection{Effective four-dimensional action} \label{sec:eff4D}

After decomposing the tensors over the extra dimensions, we can insert the decomposition of the $D$-dimensional Ricci scalar, Eq.~\eqref{eq:ricci_4+d}, into action~\eqref{eq:EH_4+d}, using Eq.~\eqref{eq:decomposition}, such that we arrive at the following effective 4-dimensional action,
\begin{align}
S=&\frac{M^2_{4+d}}{2} \int \dd^4x \sqrt{-\tilde{g}}\Big[G_4(\phi,X)\tilde{R}+G_2(\phi,X)+G_3(\phi,X)\tilde{\square}\phi \nn\\  & +f_1(\phi,X)\phi^{\mu}\phi_{\mu\nu}\phi^{\nu} +f_2(\phi,X)\phi^{\mu}\phi_{\mu\alpha}\phi^{\alpha\nu}\phi_{\nu}+f_3(\phi,X)\tilde{\square}X\Big]\,,
\label{eq:effective4d}
\end{align}
where
\begin{eqnarray}
G_4(\phi,X)&=& \int \dd^dy \sqrt{\gamma(\phi,X,y)}\,, \label{eq:G4}\\
G_2(\phi,X)&=& \int \dd^dy \sqrt{\gamma(\phi,X,y)}\left(\bar{R}_d(\phi,X,y)-2\Lambda_{4+d}-2X(2B+F)\right)\nonumber\\
&=& \int \dd^dy \sqrt{\gamma}\left(\bar{R}_d-2\Lambda_{4+d}-X\left(2\gamma^{ab}\gamma_{ab,\phi\phi}-\frac{3}{2}\gamma^{ab}\gamma^{cd}\gamma_{bc,\phi}\gamma_{ad,\phi}+\frac{1}{2}\gamma_{\phi}^2\right)\right)\,,\\
G_3(\phi,X)&=&-\int \dd^dy\sqrt{\gamma}\gamma^{ab}\gamma_{ba,\phi}\,,\\
f_1(\phi,X)&=& \int \dd^dy \sqrt{\gamma}(2C+2\bar{C}-H)\nonumber\\
&=& \int \dd^dy \sqrt{\gamma}\left[2\gamma^{ab}\gamma_{ab,\phi,X}+\frac{1}{2}\gamma_{\phi}\gamma_X-\frac{3}{4}\gamma^{ab}\gamma^{cd}\left(\gamma_{bc,\phi}\gamma_{ad,X}+\gamma_{bc,X}\gamma_{ad,\phi}\right)\right] \,, \label{eq:f1}\\
f_2(\phi,X)&=& \int \dd^dy \sqrt{\gamma}(2D+J)\nonumber\\
&=& \int \dd^dy \sqrt{\gamma}\left[\frac{3}{4}\gamma^{ab}\gamma^{cd}\gamma_{bc,X}\gamma_{ad,X}-\gamma^{ab}\gamma_{ab,X,X}-\frac{1}{4}\gamma_X^2\right] \,, \label{eq:f2}\\
f_3(\phi,X)&=& -\int \dd^dy \sqrt{\gamma}\gamma^{ba}\gamma_{ba,X} \,. \label{eq:f3}
\end{eqnarray}
The effective 4-dimensional action in Eq.~\eqref{eq:effective4d} contains three of the four free functions of the Horndeski Lagrangian, which reads~\cite{Horndeski:1974wa,Kobayashi:2011nu}
\begin{eqnarray}\label{eq:HorndeskiLagrangian}\label{eq:Kobayashiresult}
\mathcal{L}&=&G_2(\phi,X)-G_3(\phi,X)\t{\square} \phi + G_4(\phi,X)\t{R} + G_{4X}\left[(\t{\square} \phi)^2-\phi^{\mu\nu}\phi_{\mu\nu}\right]\nn\\
&& +G_5(\phi,X) \t{G}^{\mu\nu}\phi_{\mu\nu}-\frac{G_{5X}}{6}\left[(\t{\square} \phi)^3-3\t{\square} \phi \phi^{\mu\nu}\phi_{\mu\nu} +2\phi_{\mu\nu}\phi^{\nu\lambda}\phi^{\mu}{}_{\lambda}\right] \,.
\end{eqnarray}
Eq.~\eqref{eq:effective4d} does not possess the $G_5$ and $G_{5X}$ terms, the coefficients of the contraction of the Einstein tensor with second-order derivatives of $\phi$ and the cubic combination of those that maintains second-order equations of motion.
Neither does the action contain the $G_{4X}$ term, which is the coefficient for the quadratic combination of the second-derivatives of $\phi$ that preserves second-order equations of motion.
But the action contains the additional three functions $f_1$, $f_2$, and $f_3$, which can be mapped onto the beyond-Horndeski action.

\subsection{Mapping of beyond-Horndeski terms} \label{sec:mappingbeyond}

In the following, we shall perform the mapping of the effective $4$-dimensional action arising from the integration of the $D-4$ co-dimensions over the Einstein-Hilbert action (Eq.~\eqref{eq:effective4d}) onto the beyond-Horndeski action.
In particular, we can recast the extra terms $f_1$, $f_2$, and $f_3$ in Eq.~\eqref{eq:effective4d} as terms of the beyond-Horndeski action with quadratic combinations of second-order derivatives of the scalar field $\phi$. The generic form of a beyond-Horndeski action is given by~\cite{Kobayashi:2019hrl}
\begin{equation}
S^{bH}= \int \dd^4x \sqrt{-g} \left[ G(\phi,X)R+ P(\phi,X)+ Q(\phi,X)\square \phi + \sum_{i=1}^{5} A_i(\phi,X)\mathcal{L}_i \right],
\label{eq:bh4}
\end{equation}
where
\begin{eqnarray}
\mathcal{L}_1&=& \phi_{\mu\nu}\phi^{\mu\nu},\\
\mathcal{L}_2&=& (\square \phi)^2,\\
\mathcal{L}_3&=& (\square \phi) \phi^{\mu}\phi_{\mu\nu}\phi^{\nu},\\
\mathcal{L}_4&=& \phi^{\mu}\phi_{\mu\nu}\phi^{\nu\beta}\phi_{\beta}, \label{eq:bHorndeskiL4}\\
\mathcal{L}_5&=& (\phi^{\mu}\phi_{\mu\nu}\phi^{\nu})^2.
\end{eqnarray}
For Horndeski theories $A_1=-A_2$ and $A_3=A_4=A_5=0$.
Importantly, note that the full Horndeski theory, which contains also the cubic combination of the second-order derivatives of $\phi$, the term attributed to $G_{5X}$, and the contractions of the Einstein tensor with second-derivatives of $\phi$, the term attributed to $G_5$, cannot be recovered from the action~(\ref{eq:bh4}).
But these terms are also absent in Eq.~\eqref{eq:effective4d}.
The first three terms in Eq.~\eqref{eq:effective4d} are trivially mapped onto $G$, $P$ and $Q$.
The fifth integral, over $f_2$, can directly be mapped onto $\mathcal{L}_4$.
Mapping the fourth and sixth integrals from the effective action~\eqref{eq:effective4d} onto the beyond-Horndeski form in Eq.~\eqref{eq:bh4} requires more work.

Let us first consider the fourth integral, which is given by
\begin{eqnarray}
I_4&=& \int \dd^4x \sqrt{-\tilde{g}} f_1(\phi,X)\phi^{\mu}\phi_{\mu\nu}\phi^{\nu}\nn\\
&=& -\int \dd^4x \sqrt{-\tilde{g}}\phi_{\nu} \tdelta_{\mu} (f_1\phi^{\mu}\phi^{\nu})\nn\\
&=& -\int \dd^4x \sqrt{-\tilde{g}}\left(2X^2f_{1\phi}-Xf_1\tilde{\square}\phi\right)
-\int \dd^4x \sqrt{-\tilde{g}}Xf_{1X} \phi^{\mu}\phi_{\mu\nu}\phi^{\nu} \,,
\label{eq:I4}
\end{eqnarray}
where in the second step we performed an integration by parts, and we used the fact that a total derivative does not contribute to the equation of motion if we fix appropriate boundary conditions.

For the second integral on the right-hand side of Eq.~\eqref{eq:I4}, we repeat the same step by denoting $f_1^{(2)}=-Xf_{1X}$. We can repeat $n$ integration by parts to obtain the series of terms
\be\label{eq:integral_fourth}
I_4= \int\dd^4x \sqrt{-\tilde{g}} f_1(\phi,X)\phi^{\mu}\phi_{\mu\nu}\phi^{\nu} 
= \int \dd^4x \sqrt{-\tilde{g}} \left[X\sum_{i=1}^{n}f_1^{(i)}\tilde{\square}\phi- 2X^2\sum_{i=1}^{n}f_{1\phi}^{(i)} \right] \,,
\ee
where $f_1^{(i+1)}=-Xf_{1X}^{(i)}$, $f_1^{(1)}=f_1(\phi,X)$, $f_{1\phi}^{(i)}=\frac{\partial f_1^{(i)}}{\partial \phi}$, and $f_{1X}^{(n)}=0$ such that $f_1^{(2)}=-Xf_{1X}$ and $f_1^{(3)}=Xf_{1X}+X^2f_{1XX}$. Eq.~\eqref{eq:integral_fourth} makes it clear that the fourth integral $I_4$ of the effective action contributes to $P(\phi,X)$ and $Q(\phi,X)$ of the action~(\ref{eq:bh4}).

The sixth and last integral in Eq.~\eqref{eq:effective4d} can be written as
\begin{eqnarray}
I_6&=& \int \dd^4x \sqrt{-\tilde{g}} f_3\tilde{\square} X = -\int \dd^4x \sqrt{-\tilde{g}} f_3\tdelta_{\mu}(\phi^{\mu\nu}\phi_{\nu})\nn \\
&=& \int \dd^4x \sqrt{-\tilde{g}} f_{3\phi}\phi_{\mu}\phi^{\mu\nu}\phi_{\nu}- \int \dd^4x \sqrt{-\tilde{g}} f_{3X}\phi^{\alpha}\phi_{\alpha\mu}\phi^{\mu\nu}\phi_{\nu}\,.
\label{eq:integral_six}
\end{eqnarray}
The first term on the right-hand side of Eq.~\eqref{eq:integral_six} can be recast into the form of Eq.~\eqref{eq:integral_fourth} leading to contributions to $P(\phi,X)$ and $Q(\phi,X)$. The second term belongs to $\mathcal{L}_4$ in Eq.~\eqref{eq:bHorndeskiL4}. Therefore, we note that the sixth integral of the effective 4-dimensional action contributes to $P(\phi,X)$, $Q(\phi,X)$, and $A_4(\phi,X)$ of the beyond-Horndeski action.

Finally, we obtain the complete mapping of the effective 4-dimensional action (\ref{eq:effective4d}) onto the beyond-Horndeski form in Eq.~\eqref{eq:bh4},
\begin{eqnarray}
G&=& G_4, \label{eq:mappingG4}\\
P&=& G_3+ X\sum_{i=1}^n f_1^{(i)}+ X\sum_{i=1}^m f_{3\phi}^{(i)},\\
Q&=& G_2 -2X^2\sum_{i=1}^n f_{1\phi}^{(i)} -2X^2\sum_{i=1}^m f_{3\phi\phi}^{(i)},\\
A_4&=& f_2-f_{3X},\\
A_1&=& A_2=A_3=A_5=0 \,.
\label{eq:mapping_bh4}
\end{eqnarray}
This corresponds to the following Lagrangian,
\begin{equation} \label{eq:EHtoST}
\mathcal{L}= G_4(\phi,X) \tilde{R} + P(\phi,X) \tilde{\square}\phi + Q(\phi,X) + A_4(\phi,X) \phi^{\mu}\phi_{\mu\nu}\phi^{\nu\rho}\phi_{\rho} \,.
\end{equation}
Hence, the block diagonal ansatz in Eq.~\eqref{eq:decomposition} leads the $D$-dimensional Einstein-Hilbert action to a restricted class of beyond-Horndeski theories.

As noted in Sec.~\ref{sec:eff4D}, the effective actions~\eqref{eq:EHtoST} and \eqref{eq:effective4d} do not contain the compensating $G_{4,X}$ term which guarantees that the equations of motion remain second order, and the $G_5$ and $G_{5,X}$ terms are also missing. We will explore in Sec.~\ref{sec:Gauss-Bonnet} how these terms can arise from a more general $D$-dimensional action. We note here that the theory defined by the actions~\eqref{eq:EHtoST} or \eqref{eq:effective4d} remains healthy provided it satisfies the \textit{degeneracy conditions}~\cite{Langlois:2018dxi}
\begin{equation}\label{eq:degeneracy_condition}
D_0(X)=0\,,\quad~ D_1(X)=0\,,\quad~D_2(X)=0\,,
\end{equation}
where 
\begin{eqnarray}
D_0(X)&\equiv& -4(A_1+A_2)\left[4XG\left(-A_2+XA_4+ G_X\right)-2G^2-8X^2G_X^2\right]\,,\\
D_1(X)&\equiv& 8 \left[2X^2A_2(3A_1+A_2)-G^2+4XGA_1\right]A_4 + 16X^2G(A_1+A_2)A_5 \nonumber\\
&& -16XA_2^3 -4\left(G+4XG_X+12XA_1\right)A_2^2-16\left(G+5XG_X\right)A_1A_2 \nonumber\\
&& -4X^2GA_3^2 -8X\left(3G-4XG_X\right)A_2A_3-16G_X\left(G+2XG_X\right)A_1 \nonumber\\
&& + 8GG_XA_2 -8G\left(G-XG_X\right)A_3+12 GG_X^2\,,\\
D_2(X)&\equiv& 4\left[2G^2-8XGA_1-4X^2A_2(3A_1+A_2)\right]A_5 + 4(2A_1+2XA_3+2G_X)A_2^2\nonumber\\
&& +4A_2^3+12X^2A_2A_3^2+8XGA_3^2+8(G+XG_X)A_2A_3+16G_XA_1A_2\nonumber\\
&& + 4G_X^2A_2 + 8G_X^2A_1+8GG_XA_3\,,
\end{eqnarray}    
and the coupling functions $G,A_1,A_2,A_3,A_4,A_5$ parameterise the quadratic DHOST theories~\eqref{eq:bh4}. The degeneracy conditions~\eqref{eq:degeneracy_condition} for the effective action~(\ref{eq:EHtoST}) are satisfied if $A_4 = f_2 -f_{3,X} = 3 G_{4,X}^2/2 G_4$, which can be translated into a condition on the extra dimensional metric $\gamma_{mn}$.
Thus, the degeneracy conditions on the resulting scalar-tensor theories emerge as constraints on the metric of the co-dimensions.

\subsection{Special case: luminal speed of gravitational waves}

Requiring a luminal speed of gravity $c_g=c$~\cite{Monitor:2017mdv,Lombriser:2015sxa} restricts the form of the beyond-Horndeski (or quadratic DHOST) action~\eqref{eq:bh4} to~\cite{Langlois:2018dxi}
\begin{eqnarray}
A_1&=& A_2=0 \,,\label{eq:condition1_speed}\\
A_4&=& \frac{1}{8G}\left[ 12 G_X^2-8(G-XG_X)A_3-4X^2A_3^2\right] \,, \label{eq:condition2_speed}\\
A_5&=& \frac{1}{2G}\left(-2G_X -2XA_3\right)A_3 \,,
\label{eq:condition3_speed}
\end{eqnarray}
where $G$, $P$, $Q$, and $A_3$ are free functions of both $\phi$ and $X$.
It is worth noting that the mapping of the effective 4-dimensional action onto the beyond-Horndeski action in Eqs.~\eqref{eq:mappingG4}--\eqref{eq:mapping_bh4} is consistent with the conditions~\eqref{eq:condition1_speed}--\eqref{eq:condition3_speed}.
For $\gamma_{mn,X}\neq 0$, this leads to a restriction on $\gamma_{mn}(\phi,X,y)$ such that $A_4=3G_{4X}^2/{2G_4}$, which is the same requirement for satisfying the degeneracy conditions as found in Sec.~\ref{sec:mappingbeyond}. %
Note that the same beyond Horndeski term appeared first as a result of coupling matter to an $X$-dependent conformal rescaling of the metric \cite{Zumalacarregui:2013pma}.

If we further restrict to the special case where $\gamma_{mn,X}=0$, we recover 
\be\label{eq:luminalHorndeski}
\mathcal{L} = G_4(\phi) \tilde{R} + P(\phi,X) \tilde{\square} \phi + Q(\phi, X)\,,
\ee
which corresponds to the subset of Horndeski theories for which gravitational waves propagate at the speed of light,
$G_{4X}=G_5=0$~\cite{McManus:2016kxu}.

\section{$D$-dimensional Einstein-Gauss-Bonnet action}\label{sec:Gauss-Bonnet}

In Sec.~\ref{sec:Einstein-Hilbert} we have seen that the integration of the Einstein-Hilbert action over the $d$ co-dimensions with the decomposition~\eqref{eq:decomposition} can be embedded in the 4-dimensional beyond-Horndeski action~\eqref{eq:bh4}.
We have also found that a restriction to the Horndeski terms recovers the subclass of models that exhibit a luminal propagation speed of gravitational waves.
This implies that the coefficients of the quadratic ($G_{4X}$) and cubic ($G_{5X}$) combinations of second derivatives of $\phi$ and the coefficient of the contraction of second derivatives with the Einstein tensor ($G_5$) are all vanishing.
To recover these terms and therefore the full Horndeski action, one could consider a more general metric decomposition, say a non-block diagonal metric in $\gamma_{mn}$. The $G_{4X}$, $G_5$, $G_{5X}$  terms can, however, also arise from a higher-dimensional action when including higher curvature terms. This shall be the scenario of interest in the following. In particular, it is well known that a $\phi$-dependent coupling of the 4-dimensional Gauss-Bonnet term can be mapped onto the Horndeski action with contributions to $G_{4X}$, $G_{5}$, and $G_{5X}$~\cite{Kobayashi:2011nu,Kobayashi:2019hrl,vandeBruck:2018jlz} (see Sec.~\ref{sec:GBcompactification}).
We shall hence consider the extension of the Einstein-Hilbert action with the Gauss-Bonnet term in higher dimensions
\begin{equation}
S=\frac{M^2_{4+d}}{2}\int \dd^{4+d}x \sqrt{-\det(g_{AB})} \left[R- 2\Lambda_{4+d}+ \lambda G_b \right]\,,
\label{eq:EGB_4+d}
\end{equation}
where $\lambda$ is a constant and the Gauss-Bonnet term is given by
\be 
G_b=R^2-4R_{AB}R^{AB}+R_{ABCD}R^{ABCD}\,.
\ee
In $D=4$ dimensions, the Gauss-Bonnet term can be written as a boundary term, which implies that it does not contribute to the equations of motion~\cite{Padmanabhan:2013xyr}.
In higher dimensions, it leads to second-order corrections to the equations of motion such that there are no Ostrogradsky ghosts~\cite{Padmanabhan:2013xyr}. Note that the idea of recovering covariant Galileon theories as a manifestation of Poincar\'e symmetry from the Einstein-Hilbert action with a Gauss-Bonnet term in higher dimensions was already investigated in Ref.~\cite{deRham:2010eu} following a different approach.

Adopting the line element in Eq.~\eqref{eq:decomposition}, we will now cast the action~\eqref{eq:EGB_4+d} into an effective 4-dimensional scalar-tensor theory.
Note that a similar procedure was recently carried out for a metric decomposition of the form $\gamma_{mn}(x,y)=\phi(x)^2\hat{\gamma}_{mn}(y)$ of action~\eqref{eq:EGB_4+d}~\cite{vandeBruck:2018jlz}, using the embedding of the coupled 4-dimensional Gauss-Bonnet term in Horndeski theory~\cite{Kobayashi:2011nu} (see Sec.~\ref{sec:GBcompactification}).
With the adoption of the line element in Eq.~\eqref{eq:decomposition} we generalise this approach to a broader range of metric decompositions with a scalar field, recovering this special scenario for $\gamma_{mn}(\phi(x),X(x),y) = \phi(x)^2\hat{\gamma}_{mn}(y)$ (see Sec.~\ref{sec:specialcase}).
In particular, our generalisation allows for an $X$ dependence of $\gamma_{mn}$, which introduces beyond-Horndeski, DHOST, and beyond-DHOST terms.

\subsection{Tensor decomposition}

Separating out the Gauss-Bonnet contribution into terms of $\t{g}_{\mu\nu}$, we find
\begin{align}
G_b= & \left(\tilde{R}^2-4\tilde{R}_{\alpha \beta}\tilde{R}^{\alpha \beta}+\tilde{R}_{\alpha\beta \mu\nu}\tilde{R}^{\alpha \beta \mu\nu}\right)+ c_1\tilde{R}+c_2\tilde{G}^{\mu\nu}\phi_{\mu\nu}\nonumber\\ 
&+ c_3 +c_4 \tilde{\square} \phi + c_5\phi^{\mu}\phi_{\mu\nu}\phi^{\nu}+c_6\left(\tilde{\square}\phi\right)^2+ c_7\phi^{\alpha\beta}\phi_{\alpha\beta} + c_8 \left(\tilde{\square}\phi\right)\phi^{\mu}\phi_{\mu\nu}\phi^{\nu} \nonumber \\
 &+ c_9\phi^{\mu}\phi_{\mu\alpha}\phi^{\alpha\nu}\phi_{\nu} + c_{10}\left(\phi^{\mu}\phi_{\mu\nu}\phi^{\nu}\right)^2 +  c_{11}\tilde{\square}X +c_{12}\phi^{\alpha}X_{\alpha\beta}\phi^{\beta} + c_{13} \tilde{R}_{\alpha\beta} \phi^{\alpha}\phi^{\beta} \nonumber\\
  &+ c_{14} \phi^{\mu}\phi_{\mu\alpha}\phi^{\alpha\beta}\phi_{\beta\nu}\phi^{\nu}+c_{15}(\phi^{\alpha}\phi_{\alpha\beta}\phi^{\beta})\phi^{\mu}\phi_{\mu\nu}\phi^{\nu\rho}\phi_{\rho}+ c_{16}(\tilde{\square}\phi)\phi^{\mu}\phi_{\mu\alpha}\phi^{\alpha\nu}\phi_{\nu} \nn\\
  &+\left[ c_{17}\tilde{R}\phi^{\mu}\phi_{\mu\nu}\phi^{\nu} + c_{18} \tilde{R}\phi^{\mu}\phi_{\mu\alpha}\phi^{\alpha\nu}\phi_{\nu}
  + c_{19}\tilde{R}\tilde{\square}X 
 + c_{20}(\tilde{\square}\phi)\tilde{\square}X + c_{21}\left(\tilde{\square}X\right)\phi^{\mu}\phi_{\mu\nu}\phi^{\nu} \right.\nonumber \\
 &\left.+ c_{22}\left(\phi^{\mu}\phi_{\mu\alpha}\phi^{\alpha\nu}\phi_{\nu}\right)^2 + c_{23}\left(\tilde{\square}X\right)\phi^{\mu}\phi_{\mu\alpha}\phi^{\alpha\nu}\phi_{\nu} 
  + c_{24}\left(\tilde{\square}X\right)^2 + c_{25}\tilde{R}_{\alpha\beta}\phi^{\alpha}\phi^{\beta\mu}\phi_{\mu} \right.\nonumber\\
  &\left.+ c_{26}\tilde{R}_{\alpha\beta}\phi^{\alpha \mu}\phi^{\beta\nu}\phi_{\mu}\phi_{\nu} + c_{27}\tilde{R}_{\alpha\beta}X^{\beta\alpha} + c_{28} \phi_{\alpha\beta}X^{\alpha\beta}
 + c_{29}\phi_{\alpha} X^{\alpha\beta}\phi_{\beta\mu}\phi^{\mu} \right.\nonumber \\
 &\left. +c_{30} \phi^{\mu}\phi_{\mu\alpha}X^{\alpha\beta}\phi_{\beta\nu}\phi^{\nu} + c_{31} X_{\alpha\beta}X^{\alpha\beta} \right]\,,
\label{eq:GB_compact}
\end{align} 
where the $c_{i}(\phi,X,y)$ ($i=1,2,...,31$) are functions of the co-dimensional metric $\gamma_{mn}$.
In particular, they are explicitly independent of the $4$-dimensional metric $\t{g}_{\mu\nu}$. The explicit forms of the $c_i$ are given in Appendix~\ref{app:extratensors}.

On the right-hand side of Eq.~\eqref{eq:GB_compact}, the first term in the parenthesis, is the Gauss-Bonnet term constructed out of the effective $4$-dimensional spacetime metric. The terms with coefficients $c_i$ ($i={1,2,...,13}$) contribute to the various terms of the beyond-Horndeski action~\eqref{eq:bh4} with quadratic combinations of second-order derivatives of the scalar field $\phi$. Of these, the terms with coefficients $c_6,\,c_7,\,c_8,\,c_9,\,c_{10}$ are directly related to the beyond-Horndeski terms given by the $\mathcal{L}_i$ of Eq.~\eqref{eq:bh4}. 
One can furthermore show that the terms with coefficients $c_{11},\,c_{12},\,c_{13}$ can also be recast into the form of these Lagrangians, by integrating by parts and making use of the contracted Bianchi identity. The terms with the coefficients $c_{14}$, $c_{15}$, and $c_{16}$ contain cubic combinations of second-order derivatives of $\phi$. These three terms belong to the DHOST action~\cite{Langlois:2018dxi}. All other remaining terms inside the square bracket are beyond DHOST. When all these terms are inserted into Eq.~\eqref{eq:EGB_4+d}, after integration over all extra spatial $d$-dimensions, we are left with the coefficients 
\begin{equation}
C_i(\phi,X)= \int \dd^dy \sqrt{\gamma}\, c_i(\phi,X,y) \,.
\label{eq:C_i}
\end{equation}

\subsection{Higher-order derivatives}\label{sec:GBcompactification}

The contribution of the $4$-dimensional Gauss-Bonnet term to the action~\eqref{eq:EGB_4+d} is given by
\begin{equation}
S \supset \lambda \int \dd^4x \sqrt{-\tilde{g}}\, G_4(\phi,X)\left(\tilde{R}^2-4\tilde{R}_{\alpha \beta}\tilde{R}^{\alpha \beta}+\tilde{R}_{\alpha\beta \mu\nu}\tilde{R}^{\alpha \beta \mu\nu}\right) \,, 
\label{eq:GB4_action}
\end{equation}
where $G_4(\phi,X)$ represents the volume of the extra dimensions specified in Eq.~\eqref{eq:G4}.
In the following, we map out this integral to the beyond Horndeski and DHOST theories, and we also identify new terms that go beyond that.

For this purpose, let us consider the following result obtained in Ref.~\cite{Kobayashi:2011nu}. A scalar field coupled to the Gauss-Bonnet term as
\begin{equation}
\int \dd^4x \sqrt{-\t{g}} \xi(\phi)\left(\t{R}^2-4\t{R}_{\mu\nu}\t{R}^{\mu\nu}+\t{R}_{\mu\nu\rho\sigma}\t{R}^{\mu\nu\rho\sigma}\right) \,,
\label{eq:scalar_GB}
\end{equation}
can be embedded in the Horndeski Lagrangian~\eqref{eq:HorndeskiLagrangian} with
\begin{eqnarray}
G_2&=& 8 \xi^{(4)}X^2\left(3- \ln X\right)\,,\label{eq:Hfunc1}\\
G_3&=& 4\xi^{(3)}X\left(7- 3\ln X\right)\,,\\
G_4&=& 4\xi^{(2)}X\left(2- \ln X\right)\,\\
G_5&=& -4 \xi^{(1)}\ln X\,,
\label{eq:Hfunc4}
\end{eqnarray}
where $\xi^{(n)}= \frac{\partial^n \xi}{\partial \phi^n}$.
This result was used in Ref.~\cite{vandeBruck:2018jlz} to map the decomposition $\gamma_{mn}(x,y)=\phi(x)^2\hat{\gamma}_{mn}(y)$ of the action~\eqref{eq:EGB_4+d} onto Horndeski theory.
However, in our case, we cannot directly apply these expressions as the function multiplying the Gauss-Bonnet term in Eq.~\eqref{eq:GB4_action} depends not only on $\phi$ but also on $X$.

To circumvent this problem, we use the following strategy:
\begin{enumerate}
\item[(i)] We use the freedom to redefine $\xi(\phi)=\phi$ in Eq.~\eqref{eq:scalar_GB}.
\item[(ii)] We introduce an intermediate scalar field $\phi\equiv\phi(\phi',X')$, where $X'=-\frac{1}{2}\phi'^{\alpha}\phi'_{\alpha}$.
\item[(iii)] We rewrite the final result as $\xi\rightarrow G_4$, $\phi'\rightarrow \phi$, and $X'\rightarrow X$.
\end{enumerate}
Following this strategy, (i) setting $\xi(\phi)=\phi$ in Eq.~(\ref{eq:scalar_GB}), we get $\xi^{(1)}=1$, $\xi^{(2)}=\xi^{(3)}=\xi^{(4)}=0$.
Then, from Eq.~\eqref{eq:Hfunc1}--\eqref{eq:Hfunc4}, we obtain $G_5=-4\ln X$ and $G_{5X}=-\frac{4}{X}$ as well as $G_2=G_3=G_4=0$.
Next, (ii) we introduce an intermediate scalar field $\phi\equiv\phi(\phi',X')$. A derivative acting on the scalar field then implies
\begin{eqnarray}
\nabla_{\mu}\phi = \frac{\partial \phi}{\partial \phi'}\phi'_{\mu}+\frac{\partial \phi}{\partial X'} X'_{\mu}
= \xi_{\phi'}\phi'_{\mu}- \xi_{X'} \phi'_{\mu\alpha}\phi'^{\alpha}\,,
\label{eq:trans_delphi}
\end{eqnarray}
where, in the last step, we use $\phi=\xi$ and, hence, $\frac{\partial \phi}{\partial \phi'}=\frac{\partial \xi}{\partial \phi'}=\xi_{\phi'}$ as well as, similarly, $\frac{\partial \phi}{\partial X'}=\xi_{X'}$.
With Eq.~\eqref{eq:trans_delphi} we have
\begin{align}
X=- \frac{1}{2}\t{\nabla}^{\mu}\phi\t{\nabla}_{\mu}\phi
= \xi_{\phi'}^2 X' +\xi_{\phi'}\xi_{X'}\phi'^{\mu}\phi'_{\mu\alpha}\phi'^{\alpha}-\frac{1}{2}\xi_{X'}^2 \phi'_{\alpha}\phi'^{\alpha\mu}\phi'_{\mu\beta}\phi'^{\beta} \label{eq:trans_X} \,.
\end{align}
Note that by setting $\phi=\phi(\phi')=\xi(\phi')$ and $X=\xi_{\phi'}^2X'$, we recover again the full expressions of the mapping in Eqs.~(\ref{eq:Hfunc1})--(\ref{eq:Hfunc4}) in terms of the primed variables $\phi'$ and $X'$.
We show this explicitly in Appendix~\ref{app:mapping_primed}.
Using Eq.~\eqref{eq:trans_X}, we get
\begin{eqnarray}
G_5= -4\ln X 
= -4 \ln \left( \xi_{\phi'}^2 X'\right)- 4\ln \left[1+ \frac{\xi_{X'}}{X'\xi_{\phi'}}\phi'^{\mu}\phi'_{\mu\alpha}\phi'^{\alpha}- \frac{\xi_{X'}^2}{2X'\xi_{\phi'}^2}\phi'_{\alpha}\phi'^{\alpha\mu}\phi'_{\mu\beta}\phi'^{\beta}\right]
\label{eq:trans_G5}
\end{eqnarray} 
and
\begin{eqnarray}
G_{5X}= -\frac{4}{X} 
= -\frac{4}{\xi_{\phi'}^2X'}\left[1+\frac{\xi_{X'}}{X'\xi_{\phi'}}\phi'^{\mu}\phi'_{\mu\alpha}\phi'^{\alpha}-\frac{\xi_{X'}^2}{2X'\xi_{\phi'}^2}\phi'_{\alpha}\phi'^{\alpha\mu}\phi'_{\mu\beta}\phi'^{\beta}\right]^{-1}\,.
\label{eq:trans_G5X}
\end{eqnarray}
Next, we compute 
\begin{eqnarray}
\phi_{\nu\mu}&=& \t{\nabla}_{\nu}\t{\nabla}_{\mu}\phi \nn\\
&=& \xi^{(2)}_{\phi'}\phi'_{\nu}\phi'_{\mu}+\xi_{\phi'}\phi'_{\nu\mu}-\xi^{(2)}_{\phi'X'}\phi'_{\mu}\phi'_{\nu\alpha}\phi'^{\alpha}-\xi^{(2)}_{\phi'X'}\phi'_{\nu}\phi'_{\mu\alpha}\phi'^{\alpha}+\xi^{(2)}_{X'}\phi'_{\mu\alpha}\phi'_{\nu\beta}\phi'^{\alpha}\phi'^{\beta}\nn\\
&&-\xi_{X'}\phi'_{\mu\alpha}\phi'_{\nu}{}^{\alpha}-\xi_{X'}\phi'_{\nu\mu\alpha}\phi'^{\alpha} \, ,
\label{eq:t_phinumu}
\end{eqnarray}
where $\xi^{(2)}_{\phi'}=\frac{\partial^2 \xi}{\partial \phi'^2}$, $\xi^{(2)}_{\phi'X'}=\frac{\partial^2 \xi}{\partial \phi'\partial X'}$, and $\xi^{(2)}_{X'}=\frac{\partial^2 \xi}{\partial X'^2}$.
With Eqs.~(\ref{eq:trans_G5}) and (\ref{eq:t_phinumu}) we obtain
\begin{eqnarray}
G_5\tilde{G}^{\mu\nu}\phi_{\mu\nu}&=& -4\xi^{(2)}_{\phi'}\ln \left( \xi_{\phi'}^2 X'\right) \left[\t{R}^{\mu\nu}\phi'_{\mu}\phi'_{\nu}+X'\t{R}\right]-4\xi_{\phi'} \ln \left( \xi_{\phi'}^2 X'\right)\t{G}^{\mu\nu}\phi'_{\mu\nu}\nn\\
&& + (\text{beyond-DHOST terms}) \,,
\label{eq:G5_term_trans}
\end{eqnarray} 
where the beyond-DHOST terms contain arbitrary powers of combinations of second-order derivatives of the scalar field $\phi'$.
These are due to the second logarithmic term on the right-hand side of Eq.~\eqref{eq:trans_G5}.
They also contain higher-order derivatives of the scalar field $\phi'$.

We next compute the last term of Eq.~\eqref{eq:Kobayashiresult} for the primed variables,
\begin{eqnarray}
&&\left[(\t{\square} \phi)^3-3\t{\square} \phi \phi^{\mu\nu}\phi_{\mu\nu} +2\phi_{\mu\nu}\phi^{\nu\lambda}\phi^{\mu}{}_{\lambda}\right]\nn\\
&& = P'\phi'^{\mu}\phi'_{\mu\nu}\phi'^{\nu}+ A'_1 (\t{\square} \phi')^2 + A'_2 \phi'^{\mu\nu}\phi'_{\mu\nu}+A'_3 (\t{\square} \phi') \phi'^{\mu}\phi'_{\mu\nu}\phi'^{\nu} + A'_4 \phi'^{\mu}\phi'_{\mu\nu}\phi'^{\nu\rho}\phi'_{\rho} \nn\\
&& +A'_5 (\phi'^{\mu}\phi'_{\mu\nu}\phi'^{\nu})^2 + B'_1 (\t{\square} \phi')^3 + B'_2 (\t{\square} \phi')^2 \phi'^{\mu}\phi'_{\mu\nu}\phi'^{\nu} + B'_3 (\t{\square} \phi') \phi'^{\mu\nu}\phi'_{\mu\nu}\nn\\ 
&& + B'_4  (\t{\square} \phi') \phi'^{\mu}\phi'_{\mu\nu}\phi'^{\nu\rho}\phi'_{\rho} + B'_5 (\t{\square} \phi') (\phi'^{\mu}\phi'_{\mu\nu}\phi'^{\nu})^2 + B'_6 (\phi'^{\alpha\beta}\phi'_{\alpha\beta})( \phi'^{\mu}\phi'_{\mu\nu}\phi'^{\nu}) \nn\\
&& + B'_7 (\phi'^{\alpha}\phi'_{\alpha\beta}\phi'^{\beta\rho}\phi'_{\rho})( \phi'^{\mu}\phi'_{\mu\nu}\phi'^{\nu}) + B'_8 \phi'_{\mu\nu}\phi'^{\nu\rho}\phi'_{\rho}{}^{\mu} + B'_9 \phi'_{\mu}\phi'^{\mu\nu}\phi'_{\nu\rho}\phi'^{\rho\sigma}\phi'_{\sigma}\nn\\
&& + B'_{10} (\phi'^{\mu}\phi'_{\mu\nu}\phi'^{\nu})^3 + C' (\t{\square} \phi'_{\alpha})\phi'^{\alpha} + D' \phi'_{\mu} \phi'_{\nu} \phi'^{\nu\mu\alpha}\phi'_{\alpha} + (\text{beyond-DHOST terms}) \,,
\label{eq:G5X_terms_trans}
\end{eqnarray}
where the beyond-DHOST terms include higher powers (up to 8) of combinations of second-order derivatives of $\phi'$ and also higher-order derivative terms.
The coefficients $P'(\phi',X')$, $A'_{i}(\phi',X')$,  $B'_{i}(\phi',X')$, $C'(\phi',X')$, and $D'(\phi',X')$ in terms of $\xi(\phi',X')$ are specified in Appendix~\ref{app:extratensors}.
Using Eqs.~(\ref{eq:trans_G5X}) and (\ref{eq:G5X_terms_trans}) we compute $G_{5X}\left[(\square \phi)^3-3\square \phi \phi^{\mu\nu}\phi_{\mu\nu} +2\phi_{\mu\nu}\phi^{\nu\lambda}\phi^{\mu}{}_{\lambda}\right]$, which generates contributions to \textbf{all} DHOST terms in the primed variables. There are also beyond-DHOST terms appearing, which due to the inverse factor in $G_{5X}$ in Eq.~\eqref{eq:trans_G5X} cannot be thrown away by setting their coefficients to zero.

Finally, our procedure recovers Eq.~\eqref{eq:GB4_action} by using Eqs.~\eqref{eq:G5_term_trans} and \eqref{eq:G5X_terms_trans} and replacing $\phi'\rightarrow \phi$ and $X'\rightarrow X$ and $\xi(\phi',X')\rightarrow G_4(\phi,X)$.
Hence, we have generalised the scalar-tensor theory mapping of the 4-dimensional Gauss-Bonnet term coupled with the scalar field via a function $\xi(\phi)$~\cite{Kobayashi:2011nu} to the more general coupling $G_4(\phi,X)$.
We conclude that a general metric decomposition of $D=4+d$ dimensions of the Einstein-Gauss-Bonnet action leads to contributions to all the known terms present in DHOST as well as to higher-order derivative terms, which we dub beyond DHOST. It would be interesting to study if these terms spoil the healthiness of the theory. 
As in Sec.~\ref{sec:Einstein-Hilbert}, we expect degeneracy conditions to impose a constraint on the metric decomposition, but we leave a generic stability analysis for future work as the resulting action is fairly complicated and the comprehensive computations involved are hence beyond the scope of this paper. In Sec.~\ref{sec:specialcase}, we show that for an $X$-independent metric decomposition (i.e., $\gamma_{mn,X}=0$) the resulting theories are indeed healthy under a minimal constraint equation.

\subsection{Special case: $\gamma_{mn,X}=0$}\label{sec:specialcase}

As a special case of our result, let us briefly consider the restriction to
$\gamma_{mn,X}=0$, for which all the beyond-DHOST terms vanish.
The only non-zero coefficients are $C_1$, $C_2$, $C_3$, $C_4$, $C_5$, $C_6$, $C_7$, and $C_{13}$. The effective $4$-dimensional Lagrangian for the action~\eqref{eq:EGB_4+d} then becomes
\begin{eqnarray}
\mathcal{L}&=& \left(G_4(\phi)+\lambda C_1(\phi,X) \right)\tilde{R} + \left(P(\phi)+ \lambda C_4(\phi,X)\right) \tilde{\square} \phi+ \lambda C_5(\phi)\phi^{\mu}\phi_{\mu\nu}\phi^{\nu}\nn \\
&& + Q(\phi,X)+\lambda C_3(\phi,X)+\lambda C_{6}(\phi) \left[\left(\tilde{\square}\phi\right)^2-\phi^{\mu\nu}\phi_{\mu\nu}\right]+\lambda C_{13}(\phi) \tilde{R}_{\alpha\beta}\phi^{\alpha}\phi^{\beta} \nn\\
&& + \lambda C_2(\phi)\tilde{G}^{\mu\nu}\phi_{\mu\nu} 
+\lambda G_4(\phi) \left(\tilde{R}^2-4\tilde{R}_{\alpha \beta}\tilde{R}^{\alpha \beta}+\tilde{R}_{\alpha\beta \mu\nu}\tilde{R}^{\alpha \beta \mu\nu}\right) \,,
\label{eq:EGB_special}
\end{eqnarray}
where $G_4(\phi)$, $P(\phi)$, and $Q(\phi,X)$ are the same functions as in Eq.~\eqref{eq:mappingG4}--\eqref{eq:mapping_bh4}. $C_{1-7}$ and $ C_{13}$ are the volume integrals~\eqref{eq:C_i} of the $c_i$ coefficients specified in Eqs.~\eqref{eq:c_i_be}--\eqref{eq:c_i_ee}. While most of the terms obviously fit into a Horndeski action, we notice that in general $C_6\neq C_{1,X}$, which can be integrated as an extra assumption or as a constraint imposed on a restricted class of ans\"atze for the line element.
In the following, we assume $C_6 = C_{1,X}$ to ensure that the resulting theory is degenerate.
The last term on the right-hand side of Eq.~\eqref{eq:EGB_special} can be mapped into Horndeski theory using Eqs.~\eqref{eq:scalar_GB}-\eqref{eq:Hfunc4}.
$C_5$ and $C_{13}$ require more work for the mapping onto the Horndeski action. These can be recast into the known forms by integration by parts,
\begin{align}
\int \dd^4x  \sqrt{-\t{g}} C_5(\phi) \phi^{\mu}\phi_{\mu\nu}\phi^{\nu} = \int \dd^4x \sqrt{-\t{g}}  X C_5(\phi) \t{\square}\phi - 2\int \dd^4x \sqrt{-\t{g}} X^2C_{5,\phi}  
\end{align}
and
\begin{align}
\int  \dd^4x  \sqrt{-\t{g}} C_{13}(\phi)\t{R}_{\alpha\beta}\phi^{\alpha}\phi^{\beta}= & \int \dd^4x \sqrt{-\t{g}} C_{13} \phi^{\beta}\left[\t{\square}\phi_{\beta}- \t{\nabla}_{\beta}(\t{\square}\phi)\right]\nn\\
= & \int \dd^4x  \sqrt{-\t{g}} C_{13} \left[ (\t{\square}\phi)^2-\phi_{\alpha\beta}\phi^{\alpha\beta} \right] \nn\\ & -3\int \dd^4x \sqrt{-\t{g}}  XC_{13,\phi} \t{\square} \phi +2\int  \dd^4x \sqrt{-\t{g}} X^2 C_{13,\phi\phi} \,.
\end{align}
We note that the line element satisfying the restricting condition $\gamma_{mn,X}=0$ maps the $d$-dimensional Einstein-Gauss-Bonnet action into a Horndeski theory with free functions of $\phi$ only, while the $X$ dependence enters in specific form in those functions. For the simplest line element used in Ref.~\cite{vandeBruck:2018jlz} $\gamma_{mn}=\phi^2(x)\hat{\gamma}_{mn}(y)$, we have $C_6= C_{1,X}$ and we obtain 
\begin{align}
&C_1(\phi,X)= 2\phi^{d-2}\left(a_1+2\mathcal{V}d(d-1)X\right), \label{eq:C1_sp}\\
&C_2(\phi) = 8 d \phi^{d-1} \mathcal{V}\,,\\
&C_3(\phi,X) = \phi^{d-4} \left[a_2 + 4 a_1(d-2)(d-3)X + 4 \mathcal{V}d(d-1)(d-2)(d-3)X^2\right]\,,\\
&C_4(\phi,X) = -4a_1(d-2)\phi^{d-3} - 8 d (d-1)(d-2) \mathcal{V} X \phi^{d-3}\,,\\
&C_5 = C_{13}=0\,,\\
&C_6(\phi) = 4\mathcal{V} d (d-1) \phi^{d-2} = C_{1,X}(\phi)\,,\\
&G_4(\phi) = \phi^d \mathcal{V}\,,\\
&P(\phi) = -2d \phi^{d-1} \mathcal{V}\,,\\
&Q(\phi,X) = \phi^{d-2} \left[a_1-2\mathcal{V} d(d-1) X\right] -2\phi^d \Lambda_{4+d}\mathcal{V} \label{eq:Q_sp}\,, 
\end{align}
where
\be
\mathcal{V}= \int \dd^d y \sqrt{\hat{\gamma}}\,, \qquad a_1= \int \dd^d y \sqrt{\hat{\gamma}} \hat{R}_d \,, \qquad a_2= \int \dd^dy \sqrt{\hat{\gamma}} ( \hat{R}_d^2-4 \hat{R}_{ab}\hat{R}^{ab}+ \hat{R}_{abcd}\hat{R}^{abcd}) \nonumber
\ee
are constants with respect to the 4-dimensional spacetime metric as these are integrals of all extra dimensions $y$ and do not depend on $\phi(x^{\alpha})$ and $X(x^{\alpha})$.
Using Eqs.~(\ref{eq:C1_sp})--(\ref{eq:Q_sp}) and Eqs.~(\ref{eq:Hfunc1})--(\ref{eq:Hfunc4}) in Eq.~(\ref{eq:EGB_special}) we obtain the final form of the effective theory embedded in Horndeski gravity,
\begin{eqnarray}\label{eq:Horndeski_final_sp}
\mathcal{L}^f_H&=&G_2^f(\phi,X)-G_3^f(\phi,X)\t{\square} \phi + G_4^f(\phi,X)\t{R} + G_{4X}^f\left[(\t{\square} \phi)^2-\phi^{\mu\nu}\phi_{\mu\nu}\right]\nn\\
&& +G_5^f(\phi,X) \t{G}^{\mu\nu}\phi_{\mu\nu}-\frac{G_{5X}^f}{6}\left[(\t{\square} \phi)^3-3\t{\square} \phi \phi^{\mu\nu}\phi_{\mu\nu} +2\phi_{\mu\nu}\phi^{\nu\lambda}\phi^{\mu}{}_{\lambda}\right] \,,
\end{eqnarray} 
where
\begin{eqnarray}
 G_2^f&=& \phi^{d-2}\left[a_1-2\mathcal{V}d(d-1)X\right] -2\mathcal{V}\Lambda^{4+d}\phi^d + \lambda \phi^{d-4}\left[a_2+4a_1(d-2)(d-3) X \right. \nn \\
& & \left. + 4\mathcal{V} d(d-1)(d-2)(d-3) X^2(7-2\ln X)\right]\,,\\
G_3^f&=& 2\mathcal{V}d \phi^{d-1} + \lambda \phi^{d-3} \left[4a_1(d-2) + 12 \mathcal{V} d(d-1)(d-2)X(3-\ln X)\right]\,,\\
G_4^f&=& \mathcal{V}\phi^d + \lambda \phi^{d-2}\left[2a_1 + 4\mathcal{V}d(d-1)X(3-\ln X)\right]\,,\\
G_5^f&=& 4\lambda \mathcal{V} d \phi^{d-1}(2-\ln X)\,.
\end{eqnarray}
The interesting aspect of this restricted class is that it shows that the higher-order derivatives arise only because of the $X$ dependence of the extra-dimensional metric $\gamma_{mn}$. Without this dependence, the effective 4-dimensional theory is nothing else but a subclass of Horndeski gravity. It seems therefore that this dependence is a necessary ingredient to build DHOST theories and perhaps it can be seen as a general tool to generate higher-order derivative terms in the action.

\section{Generalisations} \label{sec:generalisations}

We briefly discuss two extensions to our results that one may consider to further generalise the scalar-tensor theories that can be obtained from integrating out the co-dimensions of the higher-dimensional theory.
These are an extension of the current Einstein-Gauss-Bonnet action to Lovelock gravity (Sec.~\ref{sec:lovelock}) and a disformal relation between the metrics of the geometric action and that defining geodesic free fall in the matter sector (Sec.~\ref{sec:disformal}).
Another interesting generalisation, which we shall however not discuss in more detail here, is the decomposition of the metric with additional fields such as vectors, tensors, or further scalars as a means to generate extended theories of gravity such as scalar-vector-tensor models.
We leave a more detailed exploration of these approaches to future work.

\subsection{Beyond Einstein-Gauss-Bonnet gravity in higher dimensions} \label{sec:lovelock}

In general, free function counting is a powerful tool to guess if two descriptions of the same theory can be equivalent. DHOSTs that are cubic in second-order derivatives are specified by a total of 19 free functions \cite{Langlois:2018dxi}. For the mapping to work, the number of extra spacetime dimensions should be high enough for the components of $\gamma_{mn}$ to contain these 19 free functions. With $d\geq6$ ($D \geq 10$), there is enough space to store $21$ independent functions of $\phi$ and $X$ in $\gamma_{mn}$. However, a careful look at Eq.~(\ref{eq:GB_compact}) reveals that a maximum of 12 free functions can be independently generated if we consider the Einstein-Gauss-Bonnet terms. This holds even if $d\geq 6$. Moreover, in the restricted case $\gamma_{mn,X}=0$, Horndeski theories appeared with free functions of $\phi$ but $X$ entered in a special form.

If the aim is to generate the full freedom of DHOST and Horndeski theories from higher dimensions, to generate more free functions, one could start from the most general gravitational theory in $D$ dimensions that leads to symmetric rank-two divergence-free equations of motion. These are Lovelock gravity theories in arbitrary $D$ dimensions~\cite{Lanczos,Lovelock,Charmousis:2014mia,Padmanabhan:2013xyr}.
For $D\leq 6$, Lovelock gravity includes only the Ricci scalar and the Gauss-Bonnet term.
But for higher dimensions ($D>6$), the Lovelock Lagrangian contains additional, higher-order terms that still maintain second-order field equations (in terms of the derivatives of the metric)~\cite{Lanczos,Lovelock,Padmanabhan:2013xyr}.
More specifically, for $D$ dimensions, the Lagrangian for Lovelock models is given by~\cite{Lanczos,Lovelock,Charmousis:2014mia,Padmanabhan:2013xyr} 
\begin{equation}
L= \sum_{m=1}^{(D-1)/2} c_m L_m \,,
\end{equation}
where the sum spans all integers $1\leq m <(D-1)/2$ and
\begin{equation}
L_m = \frac{1}{4^m} \delta^{A_1\,A_2\,...A_{2m}}_{B_1\,B_2\,....B_{2m}} \, 
R_{A_1\,A_2}{}^{B_1\,B_2}.....R_{A_{2m-1}\,A_{2m}}{}^{B_{2m-1}\,B_{2m}}\,.
\end{equation}
Here, $\delta^{A_1...A_{2m}}_{B_1....B_{2m}}= \epsilon_{B_1....B_{2m}}\epsilon^{A_1...A_{2m}}$ is the generalised Kronecker delta symbol defined in terms of the Levi-Civita symbol.
For $m=1$ and $m=2$ we get $L_1=R$ and $L_2=G_b$ (Gauss-Bonnet term) respectively. Therefore the next higher-order Lovelock term is
\begin{equation}
L_3= \frac{1}{64}\epsilon_{B_1B_2B_3B_4B_5B_6}\epsilon^{A_1A_2A_3A_4A_5A_6}
R_{A_1A_2}{}^{B_1B_2}R_{A_3A_4}{}^{B_3B_4}R_{A_5A_6}{}^{B_5B_6} \,,
\end{equation}
which is cubic in the Riemann tensor. By adding appropriate Lovelock terms in the $D$-dimensional action and following our dimensionality reduction scheme,  we  can  therefore  increase  the number  of  available  free  functions.
This may be an interesting avenue to pursue to generate the full freedom in the $X$ dependence of the currently restricted DHOST and Horndeski coefficients.

\subsection{Disformal transformation} \label{sec:disformal}

Another way to generate two further free functions of $\phi$ and $X$ is to express the effective action in terms of a disformally related metric
\begin{equation}
\hatg_{\mu\nu}=\Omega(\phi,X)\tilde{g}_{\mu\nu}+\eta(\phi,X)\partial_{\mu}\phi\partial_{\nu}\phi\,.
\label{eq:disform}
\end{equation}
This kind of transformation preserves the weak equivalence principle, classical particle trajectories and causality~\cite{Bekenstein:1992pj}.
A motivation for such a transformation may arise from the material part of the action, where matter fields may follow geodesics of $\hat{g}$ rather than $\tilde{g}$. This can, for instance, be the result of the matter fields being confined to the 4-dimensional manifold \cite{Zumalacarregui:2012us}.
The transformation would hence correspond to a recasting of the geometric part of the action into Jordan frame.
While removing the $X$ dependence in $\Omega$ and $\eta$ preserves the form of a Horndeski theory~\cite{Bettoni:2013diz}, it was already suspected that any kinetic dependence in the disformal functions may result in an equation of motion containing higher-order derivatives. The authors of Ref.~\cite{Achour:2016rkg} have shown that these general disformal transformations map one DHOST theory to another without violation of the degeneracy condition. In particular, they have shown that there are three families of DHOSTs which are invariant under generic disformal transformations. For instance, quadratic Horndeski theories belong to the so-called type-Ia family and so by applying a disformal transformation, one may generate another DHOST of the type-Ia family but not of the other two families~\cite{Achour:2016rkg}. In this way, it may be possible to map the effective 4-dimensional action to another DHOST containing two additional free functions by application of a disformal transformation.

\section{Conclusions} \label{sec:conclusions}

We have examined dimensionality reduction of a $D$-dimensional gravity with higher-order curvature terms to a 4-dimensional effective scalar-tensor theory for a very general co-dimensional metric ansatz.
We first studied the integration of the extra dimensions of the $D$-dimensional Einstein-Hilbert action with a cosmological constant and subsequently its extension with the supplement of a Gauss-Bonnet term.

For the Einstein-Hilbert action, this results in a restricted class of quadratic beyond-Horndeski theories which preserve the propagation speed of gravitational waves.
The degeneracy conditions for the effective theories emerge as a constraint equation on the extra-dimensional metric and, as a result, the theories remain healthy even without a $G_{4X}$ term. 
The further restriction to an extra-dimensional metric that is independent of the kinetic contribution of the scalar field $X$ yields a recovery of the full subset of $4$-dimensional Horndeski theories with luminal speed of gravity.

We then generalised the approach with the supplement of the higher-dimensional Gauss-Bonnet term.
We found that this term generates contributions to all the known DHOST terms that are cubic in second-order derivatives, but it also generates higher-order derivatives, which we dubbed `beyond DHOST'.
We defer the question of the healthiness of these extra terms to a Hamiltonian analysis, but since we started from a healthy theory, we expect healthy theories to be embedded in our beyond-DHOST terms that can be recovered with suitable constraint equations on the metric decomposition.
Interestingly, when the extra-dimensional metric is independent of $X$, i.e., $\gamma_{mn,X} =0$, all the DHOST and beyond-DHOST terms vanish and the surviving terms remain embedded in the full Horndeski theory under a minimal restriction of the $D$-dimensional line element ansatz.
As the simplest example of this scenario, we recovered and confirmed the results obtained by integration of higher-dimensional Einstein-Gauss-Bonnet gravity for $\gamma_{mn}=\phi^2(x)\hat{\gamma}_{mn}(y)$ performed in Ref.~\cite{vandeBruck:2018jlz}.

Importantly, while our general results exhibit an $X$ dependence of the coefficients of the higher-derivative terms, they are not arbitrary functions of $X$. We discussed several possible generalisations of our work that may restore the full generality of these functions.
For example, we have not used the most general action in $D>6$ dimensions that preserves the second-order nature of the equations of motion, which is given by Lovelock theory instead.
An extension of our work to Lovelock gravity would introduce additional terms in the effective 4-dimensional action.
Another aspect, which we have not addressed is the coupling to matter. We have considered the gravitational sector of the action alone and ignored the matter sector. 
For example, the matter species could couple non-universally to disformally related metrics, which preserve causality and Lorentz invariance. The gravitational sector when expressed in the Jordan frame of any of the species would then be generically more complicated. 
Finally, the considered line element is already very restrictive. In fact, we could have considered many other possibilities, such as introducing a new vector or tensor degree of freedom. Our choice was supported only by simplicity and by no means was it supported by exhaustive generality. The difficulty of generalising this line element analytically is obvious and comes from the computation of the determinant of a $D$ dimensional non-block diagonal metric. Analytical automatised tensor manipulation programmes could indeed simplify the analysis of this kind of generalisation.
We leave the extension of our results with these approaches to future work.

\section*{Acknowledgments}

S.J.~was supported by the Swiss Government Excellence Scholarship 2019 (Postdoctoral, No.~2019.0209) for foreign researchers offered via the Federal Commission for Scholarships (FCS) for Foreign Students.
C.D.~and L.L.~acknowledge the support by a Swiss National Science Foundation (SNSF) Professorship grant (No.~170547).
Please contact the authors for access to research materials.

\appendix

\section{Extra tensor definitions}\label{app:extratensors}

We report here the definitions of intermediate tensors that appear in the main text in Sec.~\ref{sec:Einstein-Hilbert}:
\begin{align}
A_{ab}=&\frac{1}{4}\gamma^{cd}\left(\gamma_{ac,\phi}\gamma_{bd,X}-\gamma_{bc,\phi}\gamma_{ad,X}\right) \,, \\
B_{ab}=& \frac{1}{4}\left(\gamma^{cd}\gamma_{bc,\phi}\gamma_{ad,\phi}-2\gamma_{ab,\phi\phi}\right)\,,\\
C_{ab}=& \frac{1}{4}\left( 2\gamma_{ab,\phi, X}-\gamma^{cd}\gamma_{bc,\phi}\gamma_{ad,X}\right)\,,\\
\bar{C}_{ab}=& \frac{1}{4}\left( 2\gamma_{ab,X,\phi}-\gamma^{cd}\gamma_{bc,X}\gamma_{ad,\phi}\right)\,,\\
D_{ab}=& \frac{1}{4}\left(\gamma^{cd}\gamma_{bc,X}\gamma_{ad,X}-2\gamma_{ab,X,X}\right)\,,\\ 
E_{abc}=& \frac{1}{2}\left[\bar{\nabla}_c(\gamma_{ab,\phi})-\bar{\nabla}_b(\gamma_{ac,\phi})\right]\,,\\
\bar{E}_{abc}=&- \frac{1}{2}\left[\bar{\nabla}_c(\gamma_{ab,X})-\bar{\nabla}_b(\gamma_{ac,X})\right]\,, \\
F^a{}_{bcd}=& \frac{1}{4}\gamma^{ae}\left(\gamma_{ed,\phi}\gamma_{bc,\phi}-\gamma_{ec,\phi}\gamma_{bd,\phi}\right)\,,\\
H^a{}_{bcd}=& \frac{1}{4}\gamma^{ae}\left(\gamma_{ed,\phi}\gamma_{bc,X}+\gamma_{ed,X}\gamma_{bc,\phi} -\gamma_{ec,X}\gamma_{bd,\phi}-\gamma_{ec,\phi}\gamma_{bd,X}\right)\,,\\
J^a{}_{bcd}=& \frac{1}{4}\gamma^{ae}\left(\gamma_{ed,X}\gamma_{bc,X}-\gamma_{ec,X}\gamma_{bd,X}\right)\,, \\
K^a{}_{bc}=& \frac{1}{2}\gamma^{ad}\left[\bar{\nabla}_d(\gamma_{bc,\phi})-\bar{\nabla}_b(\gamma_{dc,\phi})\right]\,,\\
\bar{K}^a{}_{bc}=& -\frac{1}{2}\gamma^{ad}\left[\bar{\nabla}_d(\gamma_{bc,X})-\bar{\nabla}_b(\gamma_{dc,X})\right]\,,
\end{align}
where
\begin{eqnarray}
\gamma_{\phi}=& \gamma^{ab}\gamma_{ba,\phi}\,,  \qquad B=& \gamma^{ab}B_{ba}\,, \qquad C=  \gamma^{ab}C_{ba}\,, \\ \qquad \gamma_{X}=& \gamma^{ab}\gamma_{ba,X}\,,  \qquad D=& \gamma^{ab}D_{ba}\,, \qquad \bar{C}=  \gamma^{ab}\bar{C}_{ba}  \,,
\end{eqnarray}
and $F_{ab}=F^c{}_{a c b}$, $H_{ab}=H^c{}_{a c b}$, $J_{ab}=J^c{}_{a c b}$, $K_a=K^b{}_{ab}$, and $\bar{K}_a=\bar{K}^b{}_{ab}$. The extra coefficients appearing in the integration of the Gauss-Bonnet term in Sec.~\ref{sec:Gauss-Bonnet} are specified by
\begin{align}
\label{eq:c_i_be}
c_1=& 2\left(\bar{R}_d-2X\left(2B+F\right)\right)\,,\\
c_2=& 4\gamma_{\phi}\,,\\
c_3=& \left(\bar{R}_d-2X\left(2B+F\right)\right)^2- 4\left(\bar{R}_{ab}-2X(B_{ab}+F_{ab})\right)\left(\bar{R}^{ab}-2X(B^{ab}+F^{ab})\right)\nonumber \\
& + 16X^2B^2 + 16 X K_aK^a+ 16X^2B_{ab}B^{ab}-4XK_{abc}K^{abc}-4XE_{abc}E^{abc}\nn \\
& +\left(\bar{R}_{abcd}-2XF_{abcd}\right)\left(\bar{R}^{abcd}-2XF^{abcd}\right)\,,\\
c_4=&-2\gamma_{\phi}\left(\bar{R}_d-2X\left(2B+F\right)\right) + 4\gamma_{ab,\phi}\left(\bar{R}^{ab}-2X(B^{ab}+F^{ab})\right)\,, \\
c_5=& 2\left(2C+2\bar{C}-H\right)\left(\bar{R}_d-2X\left(2B+F\right)\right)-8B\left(-\frac{1}{2}\gamma_{\phi}-X(C+\bar{C})\right)\nonumber\\
& -8\left(\bar{R}_{ab}-2X(B_{ab}+F_{ab})\right)\left(C^{ab}+\bar{C}^{ab}-H^{ab}\right)-16K_a\bar{K}^a \nonumber\\ 
& -16X B_{ab}\left(C^{ab}+\bar{C}^{
ab}\right) +4XF_{abcd}H^{abcd}+4K_{abc}\bar{K}^{abc}+4E_{abc}\bar{E}^{abc}\nn \\
& -2H_{abcd}\left(\bar{R}^{abcd}-2XF^{abcd}\right)-4B^{ab}\gamma_{ab,\phi}\,,\\
c_6=& \gamma_{\phi}^2-\gamma^{ac}\gamma^{bd}\gamma_{ab,\phi}\gamma_{cd,\phi}\,,\\
c_7=& -\gamma_{\phi}^2+\gamma^{ac}\gamma^{bd}\gamma_{ab,\phi}\gamma_{cd,\phi}\,,\\
c_8=& -2\gamma_{\phi}\left(2C+2\bar{C}-H\right)+4\gamma_{ab,\phi}\left(C^{ab}+\bar{C}^{ab}-H^{ab}\right)\,,\\
c_9=& 2(2D+J)\left(\bar{R}_d-2X\left(2B+F\right)\right)-8\left(-\frac{1}{2}\gamma_{\phi}(C+\bar{C})-X(C^2+\bar{C}^2)\right)\nonumber\\
& -8\left(\bar{R}_{ab}-2X(B_{ab}+F_{ab})\right)\left(D^{ab}+J^{ab}\right)-8\bar{K}_a\bar{K}^a-8XA_{ab}A^{ab}\nonumber \\
& -4\gamma_{ab,\phi}(C^{ab}+\bar{C}^{ab})
 -8XC_{ab}C^{ab}-8X\bar{C}_{ab}\bar{C}^{ab}+ 2\bar{R}_{abcd}J^{abcd}-4XF_{abcd}J^{abcd} \nonumber \\
 &+2\bar{K}_{abc}\bar{K}^{abc}+2\bar{E}_{abc}\bar{E}^{abc}\,,\\
c_{10}=& (2C+2\bar{C}-H)^2-8BD- 8C \bar{C}-4\left(C_{ab}+\bar{C}_{ab}-H_{ab}\right)\left(C^{ab}+\bar{C}^{ab}-H^{ab}\right)\nn\\
& -4A_{ab}A^{ab}+ 8 B_{ab}D^{ab} + 8 C_{ab}\bar{C}^{ab}+ H_{abcd}H^{abcd}\,,\\
c_{11}=& -2\gamma_X \left(\bar{R}_d-2X\left(2B+F\right)\right) + 4\gamma_{ab,X} \left(\bar{R}^{ab}-2X(B^{ab}+F^{ab})\right)\,,\\
c_{12}=& 4B\gamma_X -4\gamma_{ab,X}B^{ab}\,,\\
c_{13}=& -8 B\,,\\
c_{14}=& 4\gamma_{\phi}D-4\gamma_{ab,\phi}D^{ab}\,,\\
c_{15}=& 2\left(2C + 2\bar{C}- H\right)(2D+J)-8D(C+\bar{C}) \nonumber \\
&-8\left(C_{ab}+\bar{C}_{ab}-H_{ab}\right)\left(D^{ab}+J^{ab}\right) + 8\left(C_{ab}+\bar{C}_{ab}\right)D^{ab}- 2H_{abcd}J^{abcd} \,,\\
c_{16}=& -2\gamma_{\phi}(2D+J) + 4\gamma_{ab,\phi} (D^{ab}+J^{ab})\,, \\
c_{17}=&  2(2 C + 2 \bar{C}- H)\,,\\
c_{18}=&  2(2D+J)\,,\\
c_{19}=& -2\gamma_X\,,\\
c_{20}=&  2\gamma_{\phi}\gamma_{X}-2\gamma^{ac}\gamma^{bd}\gamma_{ab,\phi}\gamma_{cd,X}\,,\\
c_{21}=& -2\gamma_X (2 C + 2 \bar{C}- H) + 4 \gamma_{ab,X} (C^{ab}+\bar{C}^{ab}-H^{ab})\,,\\
c_{22}=&  (2D+J)^2 -4D^2- 4(D_{ab}+J_{ab})(D^{ab}+J^{ab})+ 4D_{ab}D^{ab}+J_{abcd}J^{abcd} \,,\\
c_{23}=& -2\gamma_X(2D+J) +4\gamma_{ab,X}(D^{ab}+J^{ab})\,,\\
c_{24}=&  \gamma_X^2-\gamma^{ac}\gamma^{bd}\gamma_{ab,X}\gamma_{cd,X}\,,\\
c_{25}=& -8 (C+\bar{C})\,,\\
c_{26}=& -8D\,,\\
c_{27}=& 4\gamma_X\,,\\
c_{28}=& -2\gamma_{\phi}\gamma_{X}+ 2\gamma^{ac}\gamma^{bd}\gamma_{ab,\phi}\gamma_{cd,X}\,,\\
c_{29}=& \gamma_X(C+\bar{C}) -4\gamma_{ab,X}(C^{ab}+\bar{C}^{ab})\,,\\
c_{30}=& 4\gamma_X D -4\gamma_{ab,X}D^{ab}\,,\\
c_{31}=& -\gamma_X^2+\gamma^{ac}\gamma^{bd}\gamma_{ab,X}\gamma_{cd,X}\,.
\label{eq:c_i_ee}
\end{align}
The coefficients arising from the integration of the Gauss-Bonnet action in Eq.~\eqref{eq:G5X_terms_trans} are given by
\begin{align}
P'=& 24X'^2(\xi^{(2)}_{\phi'})^2(\xi_{X'}-2\xi^{(2)}_{X'\phi'}) \,,\\
A'_1=& -6X'\xi^2_{\phi'}\xi^{(2)}_{\phi'} \,,\\
A'_2=& 6X'\xi^2_{\phi'}\xi^{(2)}_{\phi'}\,,\\
A'_3=&  12X'\xi^{(2)}_{\phi'}\xi^{(2)}_{X'\phi'}\xi_{\phi'}-6\xi^2_{\phi'}\xi^{(2)}_{\phi'}\,,\\
A'_4=&  6\xi^2_{\phi'}\xi^{(2)}_{\phi'}- 4X'(\xi^{(2)}_{\phi'})^2\xi_X' + 48X'^2\xi^{(2)}_{\phi'}(\xi^{(2)}_{X'\phi'})^2\,,\\
A'_5=&-32 X\xi^{(2)}_{\phi '}(\xi^{(2)}_{X'\phi'})^2+4X(\xi^{(2)}_{\phi'})^2\xi^{(2)}_{X'} +24X'\xi^{(2)}_{\phi'}\xi^{(2)}_{X'\phi'}\xi_{X'} - 4\xi_{\phi'}\xi^{(2)}_{\phi'}\xi^{(2)}_{X'\phi'}\,,\\
B'_1=& \xi^3_{\phi'}\,,\\
B'_2=& -6\xi^2_{\phi'}\xi^{(2)}_{X'\phi'} \,,\\
B'_3=& 12X'\xi_{\phi'}\xi^{(2)}_{\phi'}\xi_{X'} -3\xi^3_{\phi'}\,,\\
B'_4=& -12X'\xi_{\phi'}\xi^{(2)}_{\phi'}\xi^{(2)}_{X'} + 12 \xi^2_{\phi'}\xi^{(2)}_{X'\phi'} +6\xi_{\phi'}\xi^{(2)}_{\phi'}\xi_{X'} -12X'\xi_{\phi'}(\xi^{(2)}_{X'\phi'})^2\,,\\
B'_5=& 6 \xi_{\phi'}((\xi^{(2)}_{X'\phi'})^2-\xi^{(2)}_{\phi'}\xi^{(2)}_{X'})\,,\\
B'_6=& -24X'\xi^{(2)}_{\phi'}\xi^{(2)}_{X'\phi'}\xi_{X'} + 6\xi^{(2)}_{X'\phi'}\xi_{\phi'}^2 + 12 X'\xi^2_{X'}\xi^{(2)}_{\phi'} + 6\xi_{X'}\xi_{\phi'}\xi^{(2)}_{\phi'}\,,\\
B'_7=&  24X'\xi^{(2)}_{\phi'}\xi^{(2)}_{X'}\xi^{(2)}_{X'\phi'} -24\xi_{\phi'}(\xi^{(2)}_{X'\phi'})^2-12\xi^{(2)}_{\phi'}\xi^{(2)}_{X'\phi'}\xi_X' + 24 X(\xi^{(2)}_{X'\phi'})^3 \nn\\
& -12X'\xi_{X'}\xi_{X'}^{(2)}\xi_{\phi'}^{(2)}-6\xi_{\phi'}\xi_{\phi'}^{(2)}\xi_{X'}^{(2)}\,,\\
B'_8=& -12X'\xi_{\phi'}\xi^{(2)}_{\phi'}\xi_{X'} + 2\xi_{\phi'}^3\,,\\
B'_9=& 8X'\xi^{(2)}_{\phi'}\xi^{(2)}_{X'\phi'}\xi_{X'}+12X'\xi_{\phi'}\xi^{(2)}_{\phi'}\xi^{(2)}_{X'}- 16X'\xi_{\phi'}(\xi^{(2)}_{X'\phi'})^2 -12\xi_{\phi'}\xi^{(2)}_{\phi'}\xi_{X'} \nn \\ & -16\xi^2_{\phi'}\xi^{(2)}_{X'\phi'}\,,\\
B'_{10}=&~ 0\,,\\
C' =&  8{X'}^2\xi_{X'} (\xi^{(2)}_{\phi'})^2\,,\\
D' =&  4X' \xi_{X'} (\xi^{(2)}_{\phi'})^2\,.
\end{align}

\section{$\phi=\phi(\phi')=\xi(\phi')$} \label{app:mapping_primed}

Finally, we show the recovery of the result of Ref.~\cite{Kobayashi:2011nu} for the case of a scalar field coupling to the Gauss-Bonnet term.
This corresponds to $\phi=\phi(\phi')=\xi(\phi')$, $G_5=-8\ln \xi^{(1)}-4\ln X'$ and $G_{5X}=-\frac{4}{(\xi^{(1)})^2X'}$ where $\xi^{(1)}=\partial \xi/\partial \phi'$. Then 
\begin{align}
\int & \dd^4x \sqrt{-g} G_{5X}\left[(\square \phi)^3-3\square \phi \phi^{\mu\nu}\phi_{\mu\nu} +2\phi_{\mu\nu}\phi^{\nu\lambda}\phi^{\mu}{}_{\lambda}\right]\nn\\
= & -\int \dd^4x \sqrt{-g}\left( \frac{4\xi^{(1)}}{X'}\right)\left[(\square \phi')^3-3\square \phi' \phi'^{\mu\nu}\phi'_{\mu\nu} +2\phi'_{\mu\nu}\phi'^{\nu\lambda}\phi'^{\mu}{}_{\lambda}\right]\nn\\
& +24 \int \dd^4x \sqrt{-g} \xi^{(2)}\left[(\square \phi')^2-\phi'^{\mu\nu}\phi'_{\mu\nu}\right] + 24\int d^4x \sqrt{-g}\left(\frac{\xi^{(2)}}{X'}\right) (\square \phi') \phi'_{\mu}\phi'^{\mu\nu}\phi'_{\nu}\nn\\
& - 24 \int \dd^4x \sqrt{-g} \frac{\xi^{(2)}}{X'}\phi'_{\lambda}\phi'^{\lambda}{}_{\mu}\phi'^{\mu\nu}\phi'_{\nu} \,.
\label{eq:int_G5X}
\end{align}
The first two terms already have the correct form to be mapped onto a $G_{4X}$ or $G_{5X}$ term in Horndeski theory. We thus focus on the last two terms. By integration by parts, we obtain
\begin{align}
\int & \dd^4x \sqrt{-g}\left(\frac{\xi^{(2)}}{X'}\right) (\square \phi') \phi'_{\mu}\phi'^{\mu\nu}\phi'_{\nu}\nn \\
& = -\int \dd^4x \sqrt{-g} \xi^{(2)}\nabla^{\nu}\left(\ln X'\right)\phi'_{\nu}(\square \phi')\nn \\
& = \int  \dd^4x \sqrt{-g} \left[-2X'\ln X' \xi^{(3)}\square \phi' + \xi^{(2)}\ln X' (\square \phi')^2 + \xi^{(2)}\ln X' \phi'_{\nu}\nabla^{\nu}(\square \phi') \right]
\label{eq:int1}
\end{align}
and
\begin{align}
\int & \dd^4x \sqrt{-g} \left(\frac{\xi^{(2)}}{X'}\right)\phi'_{\lambda}\phi'^{\lambda}{}_{\mu}\phi'^{\mu\nu}\phi'_{\nu}\nn\\
= & -\int \dd^4x \sqrt{-g} \xi^{(2)}\phi'^{\mu\nu}\phi'_{\nu}\nabla_{\mu} \left(\ln X'\right)\nn\\
= & + \int \dd^4x \sqrt{-g} \left[-\xi^{(4)}X'^2\ln X' +\frac{1}{2}X'\ln X' \xi^{(3)}\square\phi' + \ln X' \xi^{(2)}\phi'^{\mu\nu}\phi'_{\mu\nu}+ \ln X' \xi^{(2)}\phi'_{\nu}\square \phi'^{\nu}\right] \,, \nn\\
\label{eq:int2}
\end{align}
where we used
\begin{equation}
 \int  \dd^4x \sqrt{-g} \ln X' \xi^{(3)} \phi'_{\mu}\phi'^{\mu\nu}\phi'_{\nu}
 = \int  \dd^4x \sqrt{-g}\left[\frac{1}{2}\xi^{(3)}X'\ln X' \square \phi' -\xi^{(4)}X'^2\ln X'\right] \,.
\end{equation}
Using Eqs.~\eqref{eq:int1} and \eqref{eq:int2} in Eq.~\eqref{eq:int_G5X}, we obtain
\begin{align}
\int & \dd^4x \sqrt{-g} G_{5X}\left[(\square \phi)^3-3\square \phi \phi^{\mu\nu}\phi_{\mu\nu} +2\phi_{\mu\nu}\phi^{\nu\lambda}\phi^{\mu}{}_{\lambda}\right]\nn\\
= & -\int \dd^4x \sqrt{-g}\left( \frac{4\xi^{(1)}}{X'}\right)\left[(\square \phi')^3-3\square \phi' \phi'^{\mu\nu}\phi'_{\mu\nu} +2\phi'_{\mu\nu}\phi'^{\nu\lambda}\phi'^{\mu}{}_{\lambda}\right]\nn\\
& + \int \dd^4x \sqrt{-g}\left[24\xi^{(2)}(1+\ln X')\left((\square \phi')^2-\phi'^{\mu\nu}\phi'_{\mu\nu}\right)-60X'\ln X' \xi^{(3)}\square \phi' \right.\nn\\
& \left. + 24 X'^2\ln X' \xi^{(4)}-24\xi^{(2)}\ln X' R_{\mu\nu}\phi'^{\mu}\phi'^{\nu}\right] \,,
\label{eq:int_G5Xf}
\end{align}
where we used $\nabla_{\nu}(\square \phi')-\square \phi'_{\nu}= -R_{\rho\nu}\phi'^{\nu} $. Next, we compute
\begin{align}
 \int &\dd^4x \sqrt{-g} G_5 G^{\mu\nu}\phi_{\mu\nu}\nn \\
=& -4\int  \dd^4x \sqrt{-g} \xi^{(1)}\ln X' G^{\mu\nu}\phi'_{\mu\nu} - 8\int d^4x \sqrt{-g} \xi^{(1)}\ln \xi^{(1)} G^{\mu\nu}\phi'_{\mu\nu}\nn\\
& -8\int  \dd^4x \sqrt{-g} \xi^{(2)}\ln \xi^{(1)} G^{\mu\nu}\phi'_{\mu}\phi'_{\nu}
-4\int  \dd^4x \sqrt{-g} \xi^{(2)}\ln X' G^{\mu\nu}\phi'_{\mu}\phi'_{\nu} \,.
\label{eq:int_G5}
\end{align}
By partial integration and with the Bianchi identities, we obtain
\begin{align}
 \int & \dd^4x \sqrt{-g} G_5 G^{\mu\nu}\phi_{\mu\nu}\nn \\
 = & -4\int  \dd^4x \sqrt{-g} \xi^{(1)}\ln X' G^{\mu\nu}\phi'_{\mu\nu}\nn\\
& + \int  \dd^4x \sqrt{-g} \left[ -4\xi^{(2)}\ln X' R^{\mu\nu}\phi'_{\mu}\phi'_{\nu} -4\xi^{(2)}X'\ln X' R + 8\xi^{(2)}X' R \right.\nn\\
& \left. + 8\xi^{(2)} \left((\square \phi')^2-\phi'^{\mu\nu}\phi'_{\mu\nu}\right) -24\xi^{(3)}X'(\square \phi') + 16 X'^2\xi^{(4)}\right].
\label{eq:int_G5f}
\end{align}

Using Eqs.~(\ref{eq:int_G5f}) and (\ref{eq:int_G5Xf}) as well as the additional identity $\int \dd^4x \sqrt{-g} \xi^{(3)} X' (2\ln X' -4)(\square \phi')= \int \dd^4x \sqrt{-g} \xi^{(4)} X'^2 (4\ln X' -8) $, we finally get
\begin{align}
\mathcal{L}= & + G_5 G^{\mu\nu}\phi_{\mu\nu}- \frac{G_{5X}}{6}\left[(\square \phi)^3-3\square \phi \phi^{\mu\nu}\phi_{\mu\nu} +2\phi_{\mu\nu}\phi^{\nu\lambda}\phi^{\mu}{}_{\lambda}\right]\nn\\
 = &+ G_2'-G_3'\square \phi' +G_4' R +G_{4X'}'\left[(\square \phi')^2-\phi'^{\mu\nu}\phi'_{\mu\nu} \right]+ G'_5 G^{\mu\nu}\phi'_{\mu\nu} \nn\\
& - \frac{G'_{5X'}}{6} \left[(\square \phi')^3-3\square \phi' \phi'^{\mu\nu}\phi'_{\mu\nu} +2\phi'_{\mu\nu}\phi'^{\nu\lambda}\phi'^{\mu}{}_{\lambda}\right] \equiv  \mathcal{L}'\,,
\end{align}
where
\begin{eqnarray}
G_2'&=& 8\xi^{(4)}X'^2\left(3-\ln X'\right) \,,\\
G_3'&=& 4\xi^{(3)}X'\left(7-3\ln X'\right)\,,\\
G_4'&=& \xi^{(2)}X'\left(2-\ln X'\right)\,,\\
G_5'&=& -\xi^{(1)}\ln X' \,,
\end{eqnarray}
and $\xi^{(n)}=\partial^{n}\xi/\partial \phi'^n $.
This recovers the result of Ref.~\cite{Kobayashi:2011nu} in terms of the primed variables.

\bibliographystyle{apsrev4-1}
\bibliography{reference_hd}

\end{document}